# Multiferroic-enabled magnetic exciton in 2D quantum entangled van der Waals antiferromagnet NiI$_2$

*Suhan Son*[1,2#], *Youjin Lee*[1,2#], *Jae Ha Kim*[3#], *Beom Hyun Kim*[8#], *Chaebin Kim*[1,2], *Woongki Na*[4], *Hwiin Ju*[5], *Sudong Park*[6,7], *Abhishek Nag*[9], *Ke-Jin Zhou*[9], *Young-Woo Son*[8], *Hyeongdo Kim*[6], *Woo-Suk Noh*[7], *Jae-Hoon Park*[6,7], *Jong Seok Lee*[5], *Hyeonsik Cheong*[4], *Jae Hoon Kim*[3*], *Je-Geun Park*[1,2*]*

[1]Center for Quantum Materials, Seoul National University, Seoul 08826, Republic of Korea

[2]Department of Physics and Astronomy, Seoul National University, Seoul 08826, Republic of Korea

[3]Department of Physics, Yonsei University, Seoul 03722, Republic of Korea

[4]Department of Physics, Sogang University, Seoul 04107, Republic of Korea

[5]Department of Physics and Photon Science, Gwangju Institute of Science and Technology (GIST), Gwangju 61005, Republic of Korea

[6]Department of Physics, Pohang University of Science and Technology, Pohang 37673, Republic of Korea

[7]MPPHC-CPM, Max Planck POSTECH/Korea Research Initiative, Pohang 37673, Republic of Korea

[8]School of Computational Sciences, Korea Institute for Advanced Study, Seoul 02455, Republic of Korea

[9]Diamond Light Source, Didcot OX11 0DE, United Kingdom

[10]XFEL Beamline Division, Pohang Accelerator Laboratory, Pohang 37673, Republic of Korea

Matter-light interaction is at the center of diverse research fields from quantum optics to condensed matter physics, opening new fields like laser physics. A magnetic exciton is one such rare example found in magnetic insulators. However, it is relatively rare to observe that external variables control matter-light interaction. Here, we report that the broken inversion symmetry of multiferroicity can act as an external knob enabling the magnetic exciton in van der Waals antiferromagnet $NiI_2$. We further discover that this magnetic exciton arises from a transition between Zhang-Rice-triplet and Zhang-Rice-singlet's fundamentally quantum entangled states. This quantum entanglement produces an ultra-sharp optical exciton peak at 1.384 eV with a 5 meV linewidth. Our work demonstrates that $NiI_2$ is two-dimensional magnetically ordered with an intrinsically quantum entangled ground state.

## 1. Introduction

Magnetism is a fundamentally quantum phenomenon. Yet, the formidable fortress of the current understanding has been built upon the semi-classical description of the magnetic moment. Any casual reading of the physical pictures about magnetism does not readily reveal its underlying quantum nature. However, it originates from spin, whose correct picture requires a fully quantum-mechanical and relativistic picture, i.e., the Dirac equation. This less clear quantum nature of magnetism is quite handy for a qualitative understanding of magnetism, as demonstrated in Néel's original description of the antiferromagnetic order (AF).[1] However, it is fair to note that a fully quantum entangled description was also proposed for an antiferromagnetic state as early as in the 1930s by none other than L. Landau.[2] Nevertheless, one does not need to bother with the full quantum description of Landau to understand most magnetic phase transitions.

$NiPS_3$ is an antiferromagnetic van der Waals material with the Neel temperature at 150 K and exhibits a zigzag magnetic order.[3] Although all its bulk properties do not give any hint of surprising quantum nature, a recent study[4] demonstrated that the ground state of $Ni^{2+}$ is a quantum entangled Zhang-Rice triplet (ZRT) state. This ZRT state arises from the entanglement between an electron state of the Ni ion and a hole state of surrounding S ions. The essential requirement for this exotic state is the very small charge transfer energy from S to Ni ions. Furthermore, it was shown that at the onset of the AF order, there is a high linearly polarized exciton transition to another quantum entangled Zhang-Rice singlet (ZRS) state with a record linewidth of 0.3 meV. There is also a hint that this exciton state enters some coherent state of itself. The following works showed several other salient features of $NiPS_3$: (a) magnetic field control of linear polarization,[5] (b) multiple bound states,[6] (c) a transient metallic antiferromagnetic phase,[7] and (d) light control of magnetic anisotropy.[8]

The dramatic demonstration in NiPS$_3$ of the quantum entangled ground state and its associated exotic exciton has opened the door to a fascinating field of light-matter interaction in fully quantum entangled states. A realistic hypothesis is that light coming from these materials is most likely to possess entangled quantum nature. Another interesting point is that all these observations were made on naturally occurring van der Waals materials, which have been demonstrated to be mechanically exfoliated down to monolayer.[9] In principle, this material's advantage allows a smooth integration of NiPS$_3$ into well-established machinery of 2d vdW materials, including heterostructure.[10]

With this observation and realization in mind, there have been active searches for new candidates. This report found that NiI$_2$ is a second example of such an entangled quantum state. NiI$_2$ is known to have two successive AF transitions: $T_{N1}$= 76 K and $T_{N2}$ = 59.5 K.[11] Only below $T_{N2}$, it develops a permanent electric polarization due to a spiral structure and becomes multiferroic. A small charge transfer energy for a transition between Ni and I atoms makes it favorable to host such an exotic state. Moreover, when it undergoes an AF ordering below 59.5 K, it develops a very narrow-linewidth excitation due to the ZRT-ZRS transition. Interestingly, this excitation appears to be directly linked with the multiferroicity of NiI$_2$. As a van der Waals magnet, it will be found an essential material for future optoelectronics.

## 2. Experimental Results

NiI$_2$ forms in the layered van der Waals structure with the space group of R$\bar{3}$. One Ni atom is surrounded by six I atoms, and this one unit of the Ni-I layer is separated along the c-axis by a weak van der Waals gap (Figure 1a). This Ni layer has a perfect triangular lattice of Ni moments, as shown in Figure 1b. When it is cooled below 100 K, it exhibits a weak but visible peak at $T_{N1}$= 76 K in its susceptibility before showing a sudden drop at $T_{N2}$=59.5 K (Figure 1e). According to neutron diffraction studies,[11] it develops a spiral phase with the propagation vector of q～(0.138, 0, 1.457) below $T_{N2}$. Because of such a noncollinear structure, the inverse Dzyaloshinskii-Moriya (DM) interaction or the spin-dependent metal-ligand hybridization gives rise to nonzero electric polarization, making it an example of multiferroic with the electric polarization of $P = \sim 125\ \mu C/m^2$.[12] Our second-harmonic generation measurements taken on one of many samples used for this study demonstrate the broken inversion symmetry in Figure 1f, consistent with recent reports.[13]

Figure 2a presents the absorption coefficient of single-crystal NiI$_2$ in the near-infrared and visible ranges as a function of temperature. In Figure S1, we show a sharp absorption edge around 1.6 eV and two absorption peaks at 1.51 and 1.396 eV at 2 K. It was previously reported that the spectral feature with an onset at around 1.6 eV is the charge-transfer gap transition, and the absorption peak at 1.51 eV is assigned to the $^3A_{2g} \rightarrow {}^3T_{1g}$ transition.[14] The smaller peak at 1.396 eV is assigned to the two-magnon side-band absorption, similar to the case reported in ref.[4]

However, a very sharp exciton peak, not reported before, appears at 1.384 eV (Figure S1 and Figure 2a). This peak has a full width at a half maximum of just 5.34 meV at 2 K (Figure 2b). We calculated the second derivative of the absorption coefficient with respect to energy to examine more carefully the temperature-dependent behavior of the exciton peak at 1.384 eV. (Figure 2c). The exciton peak redshifts and broadens as the temperature increases. In Figure 2d, we tracked the value of $d^2\alpha/dE^2$ at E = 1.3864 eV, i.e., at the mid-position of the exciton peak (1.384 eV) and the two-magnon side-band peak (1.396 eV). As the temperature increases, the height of the exciton peak weakens so that the local minimum at this mid-position essentially vanishes while exhibiting zero curvature $d^2\alpha/dE^2 \rightarrow 0$. The exciton peak appears below $T_{N2}$=59.5 K and is connected with the ferroelectric spiral (proper screw) formed below $T_{N2}$. This behavior implies that our presumed exciton peak cannot be assigned to a simple d-d transition. We note that a similar ultra-narrow exciton feature appears below $T_N$=150 K in NiPS$_3$ as the material goes into the zigzag antiferromagnetic order.[4] A clear difference is that, unlike NiPS$_3$, this exciton is dark in the Raman and PL spectra of NiI$_2$ (see Figure S3). The usual excitonic PL should be weak since the energy band gap is indirect.[15] However, our exciton is local, so the band gap's nature may be irrelevant here.

When we measured the angular dependency of the exciton at 1.384 eV in the geometry shown in Figure 3a, it shows explicit angular dependency on the polarization of light (Figure 3b and 3c). When NiI$_2$ enters the ferroelectric spiral phase at $T_{N2}$=59.5 K, it forms multiferroic domains in 6 directions,[12] and the most dominant domain's direction determines the angular variation of the exciton peak. This dominant domain direction does not match with a particular crystal axis of NiI$_2$ and is spontaneously fixed when NiI$_2$ enters the ferroelectric spiral phase. The angle dependency of the exciton, defined as an eccentricity of the data, is essentially independent of temperature in the low-temperature multiferroic phase and only disappears above $T_{N2}$ (Figure 3d).

In Figure 3e, three absorption peaks at 1.384 eV(exciton), 1.396 eV(two-magnon sideband) and 1.508 eV(d-d transition) are compared (Individual peaks are identified in Figure 3). The absorption by the d-d transition at 1.508 eV has no angular dependency since it is not related to the multiferroic domains. The two-magnon side-band absorption at 1.396 eV has a very small eccentricity. This is not due to the multiferroic domain effect but due to a small influence of the anisotropic effect of the nearby exciton. The angular dependency of the exciton at 1.384 eV is strong evidence of its association with the ferroelectric spiral phase.

To determine the electronic state of $Ni^{2+}$, we have carried out X-ray absorption measurements at Ni L-edges. As shown in Figure 4a, there are two clear peaks of $L_2$ and $L_3$ edges located at 850 and 870 eV, respectively. Our cluster calculation shows that these XAS data are consistent with a slightly positive charge transfer energy of $\Delta$ = 1.498 eV (see the Method and the Supporting Information). We note that this value is slightly bigger than $NiPS_3$[4] but significantly smaller than Ni oxides.[16]

We also measured the energy loss spectrum at the $L_3$ edge by using the resonant inelastic X-ray spectroscopy technique (RIXS). At the base temperature of 12.5 K, there are several clear peaks at 1 and 1.5 eV. By increasing the temperature above $T_{N1}$ and $T_{N2}$, the 1 eV peak does not show much change, while there is a visible change to the 1.5 eV peak (see Figures 4b & 4c). However, it is noticeable that the 1.5 eV peak has a well-separated two-peak structure in the ordered phase. All the peaks have a visible incident energy dependence (see Figure 4d). It is particularly noticeable that the lower peak of the 1.5 eV is only resonant at the incident energy of 850.7 eV.

## 3. Theoretical Analysis

To simulate the RIXS data, we carried out many-body calculations using the Ni-I cluster similarly to our analysis done for $NiPS_3$.[4] Physical parameters for the calculation are presented in Table 1. The spin-orbit coupling (SOC) strength of Ni $2p$ core orbitals ($\lambda_c$) was selected as the atomic value of $Ni^{2+}$ ion. We assumed the Slater-Condon parameters between Ni $3d$ orbitals ($F_{dd}^2$ and $F_{dd}^4$) and between Ni $3d$ valence and $2p$ core orbitals ($F_{pd}^2, G_{pd}^1, G_{pd}^3$) to be about 80% of the Hartree-Fock values of $Ni^{2+}$ ion.[17] $U_{dd}$ and $U_{pd}$ are averaged Coulomb interactions between $3d$ valence orbitals and between $3d$ valence and $2p$ core orbitals, respectively. We set $U_{dd} - U_{pd} = -1.4$ eV because $U_{dd} - U_{pd}$ is conventionally about $-1 \sim -2$ eV.[17] Other parameters were set to fit both XAS and RIXS spectra well. The atomic charge transfer energy $\underline{\Delta}$ is defined as the energy difference between the center of energies of $d^8$ and $d^9\underline{L}^1$ multiplets. $\underline{\Delta}$ does not solely determine the effective charge transfer energy $\Delta$ ($\Delta'$) defined as the energy difference between the lowest states of $d^8$ ($d^9\underline{c}^1$) and $d^9\underline{L}^1$ ($d^{10}\underline{c}^1\underline{L}^1$) configurations. They are given as $\Delta = \underline{\Delta} + \frac{58}{441}F_{dd}^2 - \frac{5}{441}F_{dd}^4 + 6Dq$ and $\Delta' = \underline{\Delta} + U_{dd} - U_{pd} + 6Dq$.[18] In our choice of the parameters, $\Delta$ has a positive value of 1.498 eV, slightly larger than NiPS₃,[4] but $\Delta'$ has a negative value of $-1.12$ eV. Negative $\Delta'$ implies that the main XAS peak is dominantly attributed to the states with $d^{10}\underline{c}^1\underline{L}^1$ configuration, while the states with $d^9\underline{c}^1$ configuration, which determines the conventional XAS shape of Ni²⁺ systems, bears the subpeak structure (Figure S3).

The multiplet structure of initial states is directly affected by the effective charge transfer energy Δ. It can be estimated as 1.498 eV in our selection. Its value is slightly larger than expected one of NiPS₃ in ref.[4] ($\sim 0.95$ eV). The local electronic properties of NiI₂ would be qualitatively the same as those of NiPS₃. Indeed, the calculated multiplet structure is similar to NiPS₃, as shown in Figure S5. The coupling between the multiplet states with a $d^8$ configuration and charge transfer states with $t_{2g}^6 e_g^3 \underline{L}^1$ and $t_{2g}^5 e_g^4 \underline{L}^1$ configurations located at about 1.498 eV and 2.298 eV, respectively, derive the multiplet structure as shown in Figure S4. This multiplet structure is well reflected in the theoretical RIXS spectra.

The ground states with the $^3A_{2g}$ symmetry have a strong coupling with $t_{2g}^6 e_g^2 \underline{L}^1$ states via the σ-type bond. The contribution of $t_{2g}^6 e_g^2 \underline{L}^1$ states is about 37 %. Thus, non-negligible self-doped I $p$ holes are expected to be populated and make the ZRT-like formation with Ni $3d$ orbitals in the ground state. Excited states with the $^3T_{2g}$ and $^3T_{1g}$ symmetries are located around 1.0 eV and 1.5 eV, respectively. It implies that the peaks at around 1.0 and 1.5 eV of the optical absorption and RIXS spectra are attributed to the local $dd$ transitions. Indeed, the calculated RIXS spectra show these peaks clearly, as shown in Figure 5. Moreover, the nondegenerate state with the $^3A_{1g}$ symmetry, in which the contribution of $t_{2g}^6 e_g^2 \underline{L}^1$ states is about 71 %, appears at around 1.4 eV. This state results from the ZRS formation having following the wave function:

$$|\Psi(^1A_{1g})\rangle \approx \frac{0.835}{2}\left(\left|d_{z^2\uparrow};p_{z^2\downarrow}\right\rangle - \left|d_{z^2\downarrow};p_{z^2\uparrow}\right\rangle + \left|d_{x^2-y^2\uparrow};p_{x^2-y^2\downarrow}\right\rangle - \left|d_{x^2-y^2\downarrow};p_{x^2-y^2\uparrow}\right\rangle\right) +$$
$$\frac{0.473}{\sqrt{2}}\left(\left|d_{z^2\uparrow};d_{z^2\downarrow}\right\rangle + \left|d_{x^2-y^2\uparrow};d_{x^2-y^2\downarrow}\right\rangle\right) +$$
$$\frac{0.215}{\sqrt{2}}\left(\left|p_{z^2\uparrow};p_{z^2\downarrow}\right\rangle + \left|p_{x^2-y^2\uparrow};p_{x^2-y^2\downarrow}\right\rangle\right) + \cdots,$$

where $d_{\mu\sigma}$ and $p_{\nu\sigma}$ ($\mu,\nu \in \{z^2, x^2-y^2\}$) are hole states of Ni $3d$ orbital μ and spin σ, and I $5p$ orbital ν and spin σ, respectively. $p_{z^2}$ and $p_{x^2-y^2}$ are $p$ orbitals, which have the same symmetry as $d_{z^2}$ and $d_{x^2-y^2}$ orbitals, respectively, as following

$$p_{z^2} = \frac{1}{\sqrt{12}}(-p_{1x} - p_{2y} + p_{4x} + p_{5y} + 2p_{3z} - 2p_{6z}),$$

$$p_{x^2-y^2} = \frac{1}{2}(p_{1x} - p_{2y} - p_{4x} + p_{5y}),$$

where $p_{j\alpha}$ ($\alpha \in \{x, y, z\}$) is the $p$ orbital at the $j$-th I atom. In our optical measurements (Figures 2 and 3), the exciton-like peak with ultra-narrow width and an additional peak appears around that region. These two peaks would be directly attributed to the ZRS state or its side-band state coupled with other types of excitations like phonon, magnon, or double-magnon.

We could reproduce the energy loss spectra using the parameters shown in the SI through these many-body calculations. Our simulated data show the 1 eV peak and the two-peak structure at 1.5 eV (see Figure 4c). Furthermore, our simulation in Figure 4e for the dependency of the incident energy show remarkably similar data to the experimental data in Figure 4d. This good agreement between the experimental RIXS data and the theoretical results gives strong confidence in choosing our parameters as listed in the SI.

According to the multiplet structure and theoretical RIXS spectra, we can attribute that the RIXS peaks at 1.0 and 1.5 eV are the local $dd$ transition from the ZRT ground state with $^3A_{2g}$ symmetry to excited states with the $^3T_{2g}$ and $^3T_{1g}$ symmetries, respectively. In addition, we can conclude that the peak at 1.4 eV is the transition from the ground ZRT states to the excited ZRS state with $^1A_{1g}$ symmetry. Because the ZRS state has a different spin state compared to the ZRT ground states, the peak at 1.4 eV shows dominant intensity not at the on-resonant energy (A-cut) but at off-resonant energy (B-cut) in contrast with other peaks at 1 and 1.5 eV (see Figure 5). Its incident energy dependency supports the relevance of our interpretation on the peak at 1.4 eV.

Finally, we comment on a clear difference between NiI$_2$ and NiPS$_3$. First and foremost, NiI$_2$ has a spiral magnetic structure of a triangular lattice, while the ground state of NiPS$_3$ is a zigzag spin chain of a honeycomb lattice. With the antiferromagnetic interaction, this triangular lattice of NiI2 is geometrically frustrated with further exciting possibilities. This different magnetic and lattice structure makes the ZRT state of NiI2 a truly two-dimensional object while it is a quasi-one-dimensional for NiPS3. Another interesting difference between the two systems is the critical mechanism enabling the magnetic exciton. For NiPS$_3$, the exciton peak is enabled by the spontaneous symmetry breaking due to spin-dependent hopping. On the other hand, for NiI$_2$, the inversion symmetry is broken by the noncollinear spiral magnetic structure and thus the inverse DM interaction or the spin-dependent metal-ligand hybridization.

For instance, we tried to check whether the spin-charge coupling proposed for NiPS$_3$ [4] is viable to understand the magnetic exciton of NiI$_2$. We consider the ZRT formation in the proper screw spin order. Figures 1c and 1d depict the schematic diagram of the spin configuration of the ZRT formation in the proper screw magnetic order. In NiI$_2$, every I atom neighbors three adjacent Ni atoms. Among three neighboring Ni atoms, spins on two Ni atoms have the same orientation, and the screw modulation rotates the other. Note that the spin ordering of neighboring Ni atoms is critical in the spin-charge coupling in NiPS$_3$.[4] So, the exact mechanism for the spin-charge coupling may not seem directly applicable to NiI$_2$.

Instead, the electric polarization induced by a conventional mechanism such as the inverse DM interaction or the spin-dependent metal-ligand hybridization may play a more critical role in the spin-charge coupling. For example, if we consider the ZRS formation in the screw spin order, spin states of six ligand holes bounded with Ni d orbitals are reversed when the transition from the ZRT to ZRS states occurs shown in Figure 1d. This transition causes the spin state of two I atoms highlighted by the solid open circle in Figure 1d to be different from the remaining four ligand holes. It might enhance the two-fold charge modulation, but the inversion symmetry breaking is questionable. Consequently, the electric polarization induced by the inverse DM interaction or the spin-dependent metal-ligand hybridization could be one of the sources to make spin-charge coupling in NiI$_2$.

Conventionally, the magnetic exciton from the ZRT to ZRS states is the Frankel-type exciton with the even parity. Thus, it is optically dark in one photon process due to the spin and dipole sum rules. To be bright, both time-reversal and inversion symmetries should be broken. In the NiPS$_3$ case, the spin-charge coupling induced by the self-doped ligand holes in the zigzag order can beak both symmetries and makes it to be bright.[4] In the NiI$_2$ case, the screw spin order can also break both symmetries. Finally, the magnetic exciton is still dark.

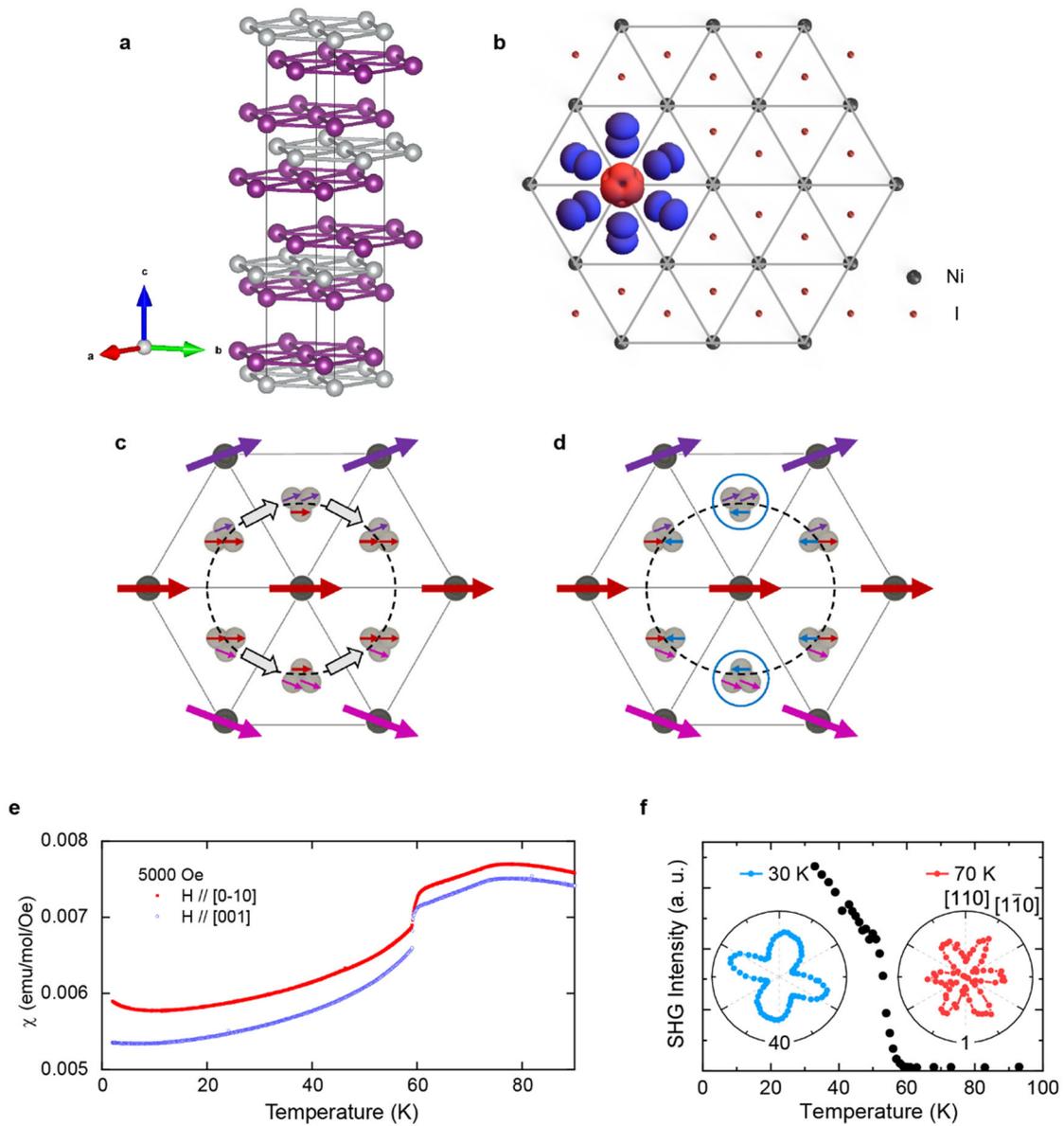

**Figure 1.** Bulk properties of NiI$_2$ and the formation of the Zhang-Rice triplet and singlet states in the proper screw spin structure. a) Crystallographic structure of NiI$_2$. b) Schematic view of Zhang-Rice triplet states. Blue and red colors illustrated the spin density difference. c) Black balls refer to Ni$^{2+}$ ions. Objects with three lobes depict holes of ligand I ions coupled to d orbitals of Ni atoms with sigma bonding. Large and small arrows represent spin moments of Ni d and I p holes, respectively. Electric polarization vectors of the ground state are highlighted with gray arrows. d) When the transition from the Zhang-Rice triplet to Zhang-Rice singlet states occurs, spin states of six ligand holes bounded with Ni d orbitals are reversed. Spin states of two I ligand holes highlighted with solid circles are expected to be different from those of the remaining four ligand holes. e) Magnetic susceptibility of NiI$_2$ single-crystal depending on the magnetic field direction. Two transitions were observed at $T_{N1}$ = 76 K and $T_{N2}$ = 59.5 K. f) SHG (Second Harmonic Generation) signals develop only below $T_{N2}$ with the insert of the angular dependence of the SHG data taken at 30 and 70 K. The directions of crystalline axes are shown in the inset.

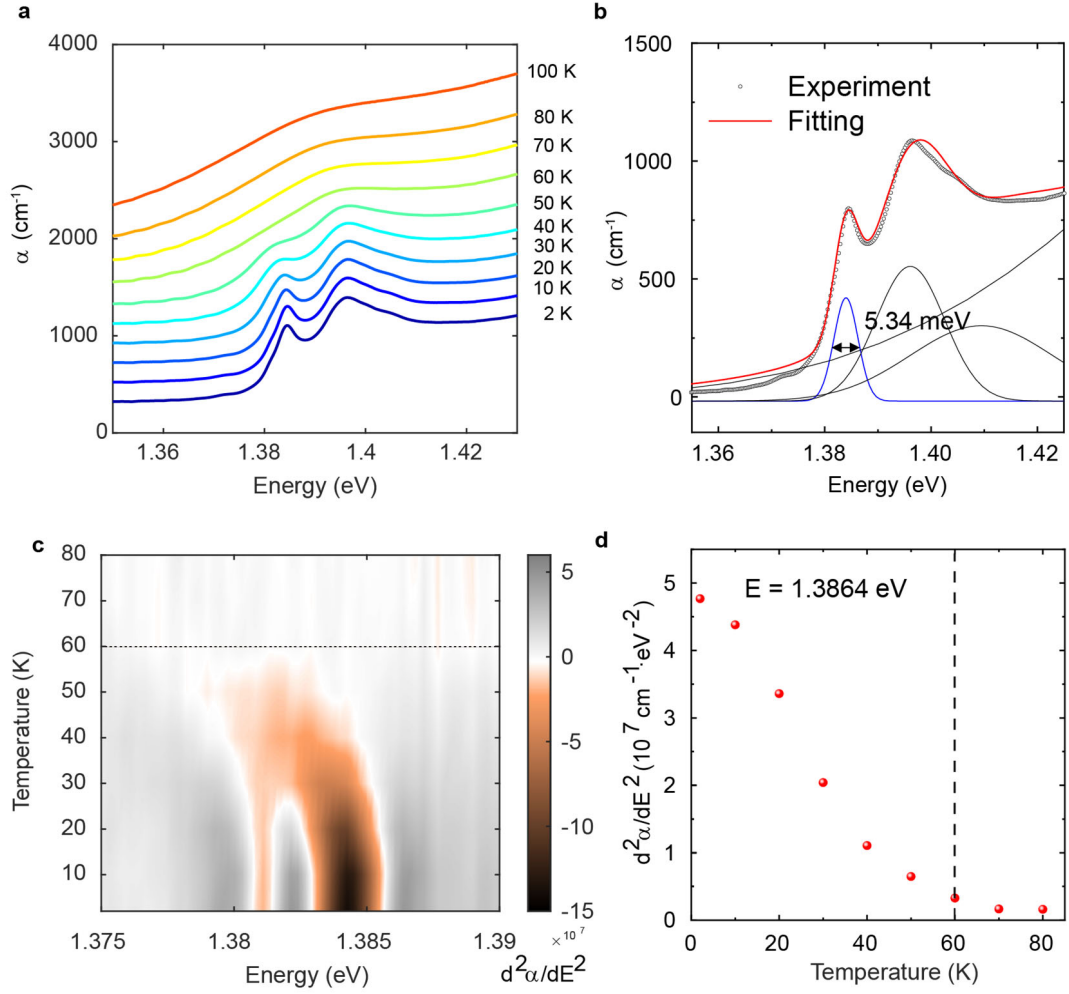

**Figure 2.** Absorption spectra of NiI$_2$ single crystal. a) The in-plane absorption coefficient of single-crystal NiI$_2$ in the ferroelectric spiral (proper screw) phase at various temperatures. b) Model fitting curves of the absorption coefficient at T = 2 K. c) Second derivative of the absorption coefficient with respect to energy. d) Value of the second derivative at E = 1.3864 eV as a function of temperature. The dotted lines in c) and d) indicate $T_{N2}$ = 59.5 K.

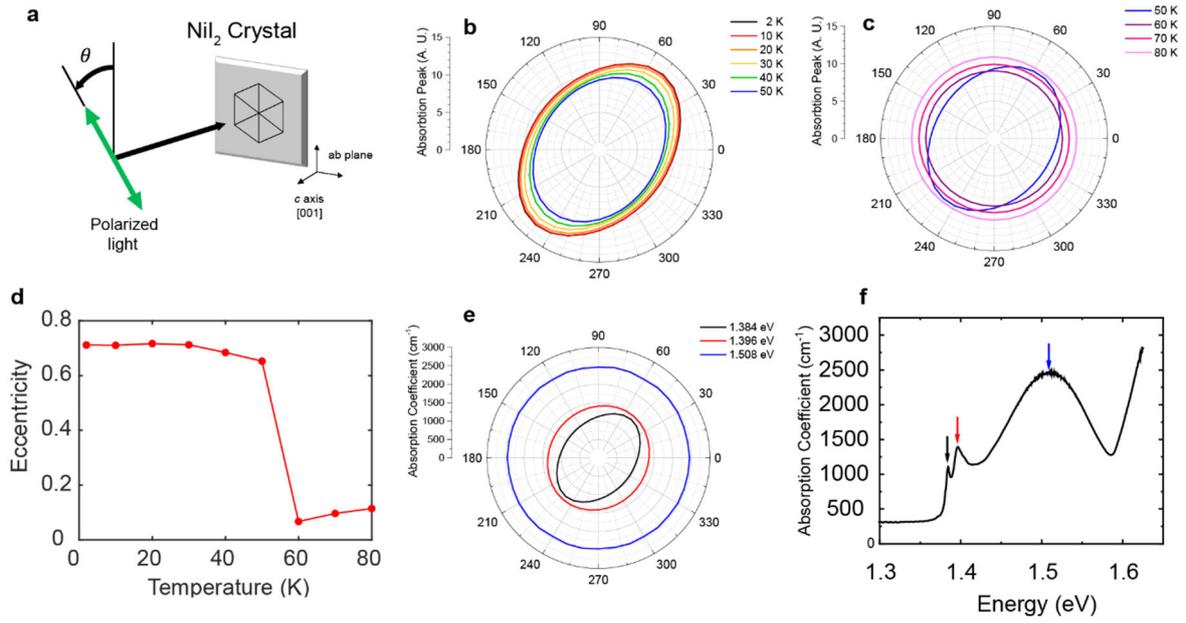

**Figure 3.** Angular dependence of the exciton peak at 1.384 eV on the polarization of light. a) Schematic of the optical spectroscopy setup. An angular plot of the peak height of exciton for b) 2 – 50 K and c) 50 – 80 K. d) Eccentricity of the angular trace of the exciton as a function of temperature. e) Angular trace plot of exciton compared with those of the nearby two absorption peaks at 2 K. f) Absorption coefficient at 2 K where the colors of arrows match those of angular traces in e).

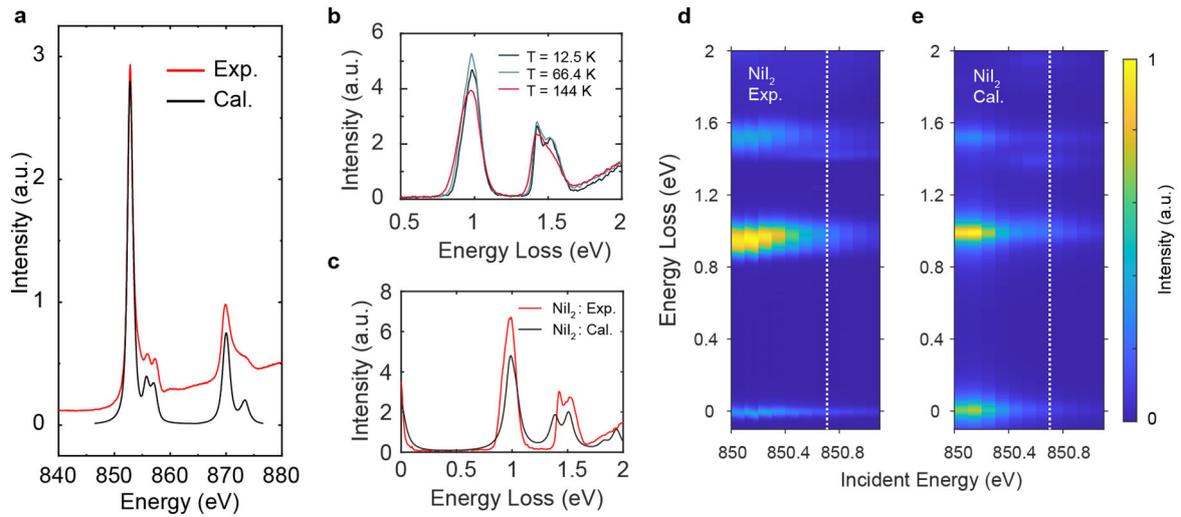

**Figure 4.** X-ray spectroscopy data. a) X-ray absorption and cluster calculation result. Parameters used for fitting were described in the main text. b) Temperature-dependent resonant inelastic X-ray scattering (RIXS) spectra. c) The energy loss spectrum of the experiment at T = 12.5 K (Red) and calculation (Black) results at the incident energy E = 850.7 eV, (dashed line in d) and e)). At the peak near 1.4 eV, the maximum intensity was observed at the given incident energy. d) Incident-energy-dependent RIXS spectra near 850.7 eV from experiments (Exp.). e) Calculated (Cal.) result based on a $NiI_6$ cluster.

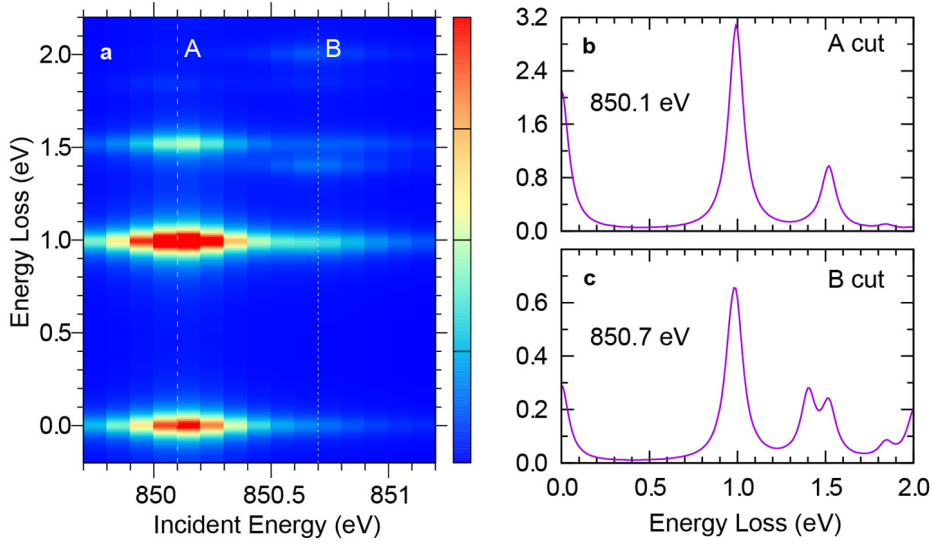

**Figure 5.** Theoretical data of RIXS. a) The theoretical Ni $L_3$-edge resonant inelastic x-ray scattering (RIXS) map. The Ni $L_3$-edge RIXS intensity as a function of the energy loss when the incident x-ray energy is b) 850.1 eV (A cut in a)) and c) 850.7 eV (B cut).

| $10Dq$ | $\underline{\Delta}$ | $U_{dd}$ | $F_{dd}^2$ | $F_{dd}^4$ | $\lambda_c$ | $U_{pd}$ | $F_{pd}^2$ | $G_{pd}^1$ | $G_{pd}^3$ | $V_{pd\sigma}$ | $V_{pd\pi}$ |
|---|---|---|---|---|---|---|---|---|---|---|---|
| 0.8 | -0.2 | 6 | 9.7872 | 6.0768 | 11.35 | 7.4 | 6.1773 | 4.6304 | 2.6334 | -0.84 | 0.5 |

**Table 1.** Physical parameters for the configuration interaction calculation. The $10Dq$ is the cubic crystal field parameter of $3d$ orbitals. $\underline{\Delta}$ is the atomic charge transfer energy defined as the energy difference between the center of energies of $d^8$ and $d^9\underline{L}^1$ multiplets. $U_{dd}$ and $U_{pd}$ are averaged Coulomb interactions between $3d$ valence orbitals and between $3d$ valence and $2p$ core orbitals, respectively. $F_{dd}^2$, $F_{dd}^4$, $F_{pd}^2$, $G_{pd}^1$, and $G_{pd}^3$ are the Slater-Condon parameters. $\lambda_c$ is the spin-orbit coupling strength of Ni $2p$ core orbitals. $V_{pd\sigma}$ and $V_{pd\pi}$ are the hopping strengths of $\sigma$- and $\pi$- type bonds between Ni $3d$ and I ligand $5p$ orbitals, respectively. Unit is eV.

## 4. Conclusion

Realizing the quantum entangled state is a pressing issue at the age of quantum information and quantum materials. We have demonstrated that NiI$_2$ has a unique combination of several situations, making it favorable for such an exotic state. Because of the weak charge transfer between Ni and I atoms, the ground state wave function of Ni$^{2+}$ has a significant electron donation from the neighboring I atoms, which then get themselves entangled with a Zhang-Rice triplet state. In addition, a novel magnetic-exciton state is enabled by multiferroicity when it undergoes a second AF transition with the broken inversion symmetry due to inverse DM interaction. Finally, the triangular lattice of Ni atoms makes the low-temperature ordered phase an ideal two-dimensional version of the quantum entangled state, thereby putting NiI$_2$ in a unique position to explore quantum entanglement in two dimensions.

**Experimental Section**

*Sample preparation and characterization*: Single crystals of NiI$_2$ were synthesized via a chemical vapor transport method. Nickel powder (99.99% Sigma-Aldrich) and crystalline iodine (99.99%, Alfa Aesar) were weighed in the stoichiometric ratio with additional 5% Iodine within an argon-filled glove box. The mixture was sealed in a quartz tube, evacuated by a rotary vane pump, and then placed in a two-zone furnace and heated to 750℃ / 720℃ over 6 hours. The typical pressure inside the quartz tube was ~1 Pa. The furnace was held for 7 days, cooled slowly to room temperature over 5 days. The crystals formed shiny gray flakes, and the typical crystal size was $5 \times 5 \times 0.1\ mm^3$.

*Bulk property measurements*: We investigated the crystallinity and the quality of NiI$_2$ single crystal by Energy-dispersive X-ray spectroscopy (EDX), X-ray diffraction (XRD), and magnetic susceptibility measurement. The sample stoichiometry was verified by a scanning electron microscope (COXEM, EM30) equipped with an EDX detector (QUANTAX 80, Bruker). The ratio of elements shows Ni:I = 1:2 within the error range. XRD measurements were carried out using an X-ray diffractometer (MiniFlex II, Rigaku) with Cu target Kα1 and Kα2 wavelengths. The temperature and magnetic field-dependent magnetization were measured by SQUID magnetometer (Quantum Design, MPMS3). Single crystals were characterized in temperature ranges from 2 to 300 K. Two successive antiferromagnetic phase transitions were observed at $T_{N1}$ = 76 K and $T_{N2}$ = 59.5 K, as reported by Karumaji et al.[12]

*XAS measurements*: The Ni $L_2$- and $L_3$-edge XAS spectrum was obtained at the 6A MPK MEXIM Beamline of PLS-II at Pohang Accelerator Laboratory. A single-crystalline sample was cleaved in-situ, and the measurements were carried out with the total electron yield at room temperature in a vacuum better than $2 \times 10^{-8}$ Torr. The photon energy resolution was set to be 0.25 eV.

*RIXS measurements*: RIXS experiments were performed following the same procedure as the previous report.[4] The measurement was carried out at the I21 beamline of the Diamond Light Source, UK. Since the sample was hygroscopic, the sample was mounted on a copper sample holder as soon as possible after breaking the quartz tube containing $NiI_2$ samples. Furthermore, the sample's surface was exfoliated right before making a vacuum by Scotch tape to obtain a fresh and flat surface. Ni $L_3$-edge was chosen by X-ray absorption spectroscopy (XAS). From the XAS measurements, we selected the principal energy (850.7 eV) to perform the RIXS experiments. The energy resolution of RIXS in this study was determined to be 36 meV. To track the temperature evolution of peaks, we changed the temperature of the sample from 12.5 to 144 K, which is far below and above critical temperatures.

*Raman and PL measurements*: A bulk single crystal of $NiI_2$ was put into an optical cryostat (Montana instruments, s50) immediately after it was taken out of the quartz tube and kept in a vacuum since the $NiI_2$ sample can be easily degraded in ambient conditions. All the measurements were carried out with the sample in a vacuum to avoid degradation during measurements. Temperature-dependent Raman and photoluminescence (PL) spectra were measured at temperatures from 4 to 293 K. The 514.5 nm (2.41 eV) line of Ar-ion laser was used as the excitation source. The laser power was kept below 100 μW to avoid damaging the samples. The laser beam was focused onto the sample by a ×40 microscope objective lens (0.6 N.A.), and the scattered light was collected and collimated by the same objective. The scattered signal was dispersed by a JobinYvon Horiba iHR550 spectrometer (300 grooves/mm for PL, 2400 grooves/mm for Raman) and detected with a liquid nitrogen-cooled back-illuminated charge-coupled-device (CCD) detector. Volume holographic filters (Optigrate) were used to clean the laser lines and reject the Rayleigh-scattered light. Finally, an analyzer was used to selectively pass scattered photons with parallel or cross-polarization for polarized optical measurements. Finally, another achromatic half-wave plate was placed in front of the entrance slit to keep the polarization direction of the signals entering the spectrometer constant with respect to the groove direction of the grating.

*Optical Absorption Measurements*: The optical absorption experiments were conducted using a CCD spectrometer (CCS175, Thorlabs) over 500-1000 nm (1.24 – 2.48 eV) and a resolution better than 2 meV. The light source is a broadband tungsten-halogen lamp covering 360 – 2600 nm. The light is guided through a polarizer to conduct a polarized transmission experiment. The experiments were conducted over the temperature range of 2 – 300 K using a He-free magneto-optic cryostat (SpectromagPT, Oxford). The sample had a thickness of about 32 μm and an area of 4 mm x 4 mm. The sample was thinned by mechanical exfoliation and mounted on an Au-coated Cu holder with a 3 mm diameter hole.

*Many-body Calculations*: To interpret the electronic excitation spectra of NiI$_2$, we performed the configuration interaction (CI) calculation of NiI$_6$ cluster incorporating all local correlation effects of Ni $3d$ valence and $2p$ core orbitals and charge transfer effect of I $5p$ orbital. For simplicity, we assumed that the ligand states of I ions have the same energy level and ignored their spin-orbit coupling (SOC) effect. Hopping integrals between the Ni $3d$ and I $5p$ orbitals were determined with the Slater–Koster theory. By taking into account all possible states with $d^8$, $d^9\underline{L}^1$, and $d^{10}\underline{L}^2$, where $\underline{L}$ refers to the ligand-hole state of I ions, we solved the CI Hamiltonian and obtained the electronic multiplet structures. To calculate the x-ray absorption spectroscopy (XAS) and resonant inelastic x-ray scattering (RIXS) spectra, we also performed the CI calculation considering all possible states with $d^9\underline{c}^1$ and $d^{10}\underline{c}^1\underline{L}^1$ configurations, where $\underline{c}$ is the $2p$ core-hole state of Ni ions. We took into account the SOC effect of Ni $2p$ core-hole states and the Coulomb interaction between the Ni $2p$ core and the $3d$ valence orbitals to describe the final states (intermediate states) of the L-edge XAS (RIXS) process. We employed the Kramers–Heisenberg equation in the dipole approximation to calculate the RIXS spectra.[19]

**Supporting Information**
Supporting Information is available from the Wiley Online Library or the author.

**Acknowledgments**

SS, YL, JHK, and BHK contributed equally to this work. We acknowledge the helpful discussion with Ki-Hoon Lee. Work at CQM and SNU was supported by the Leading Researcher Program of the National Research Foundation of Korea (Grant No. 2020R1A3B2079375). Work at Yonsei University was supported by National Research Foundation (NRF) grants funded by the Korean government (MSIT; grant 2021R1A2C3004989) and the SRC program (vdWMRC; grant 2017R1A5A1014862). Work at Sogang Univ. was supported by the National Research Foundation (NRF) grant funded by the Korean government (MSIT) (2019R1A2C3006189; 2017R1A5A1014862, SRC program: vdWMRC center). BHK and YWS at KIAS were supported by KIAS Individual Grants (CG068702), and numerical computations have been performed with the Center for Advanced Computation Linux Cluster System at KIAS. The work at POSTECH and MPK was supported by the National Research Foundation of Korea (NRF), funded by the Ministry of Science and ICT (Grant No. 2016K1A4A4A01922028 and No. 2020M3H4A2084417). The work at GIST was supported by the Ministry of Science, ICT, and Future Planning (Nos. 2015R1A5A1009962, 2018R1A2B2005331). We acknowledge Diamond Light Source for time on beamline I21.

# Supporting Information

**Optical Absorption Measurements.**

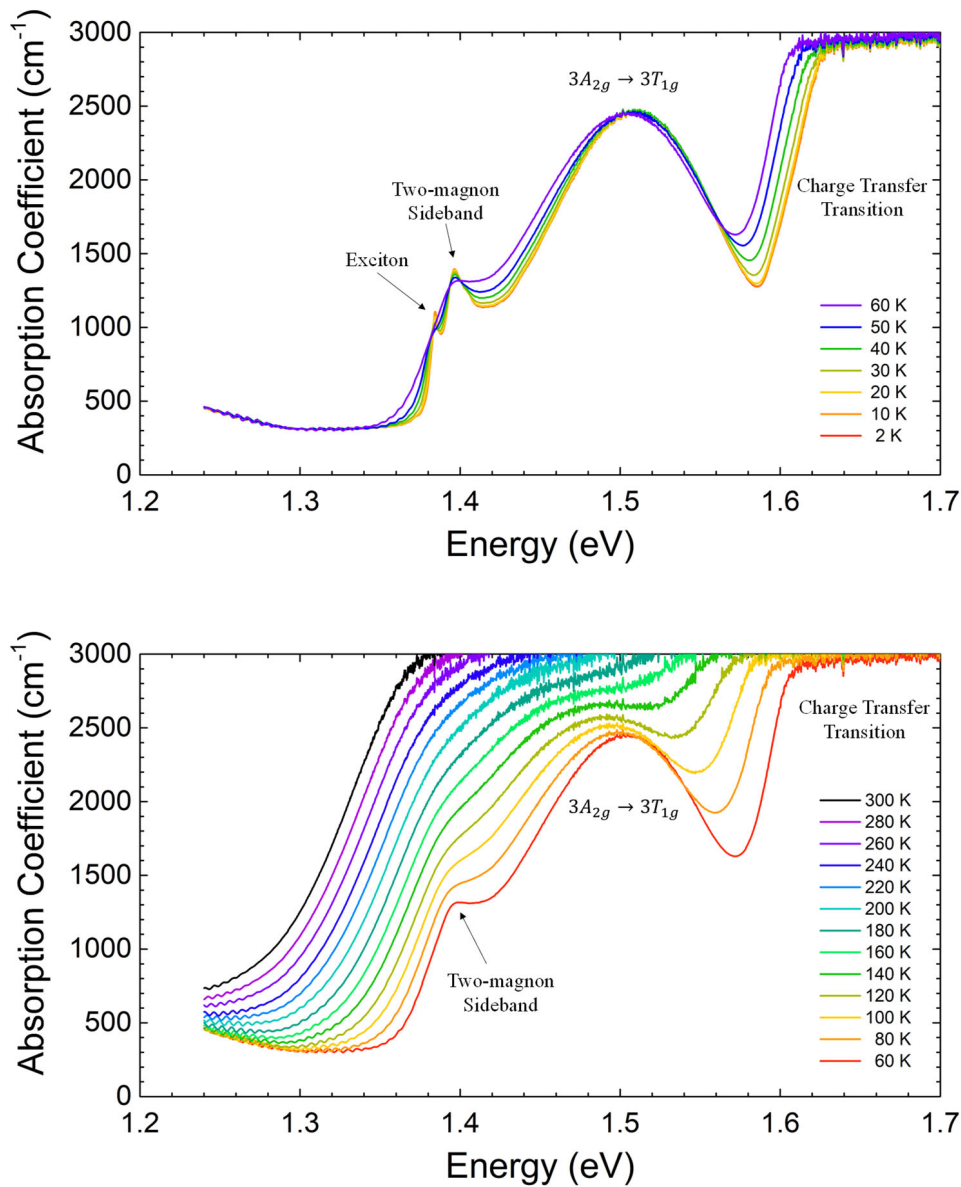

**Figure S1.** The in-plane absorption coefficient of single crystal $NiI_2$ in the ferroelectric spiral phase (top panel) and the other phases ($T > T_{N2} = 59.5$ K, bottom panel). Fig. S1 shows the temperature-dependent absorption coefficient spectra of single crystal $NiI_2$. The charge-transfer onset at 1.6 eV moves progressively towards lower energy as the temperature increases, masking the two-magnon side-band of 1.396 eV and a d-d transition of 1.508 eV. However, apart from this, we do not observe any new features developing across $T_{N1} = 76$ K.

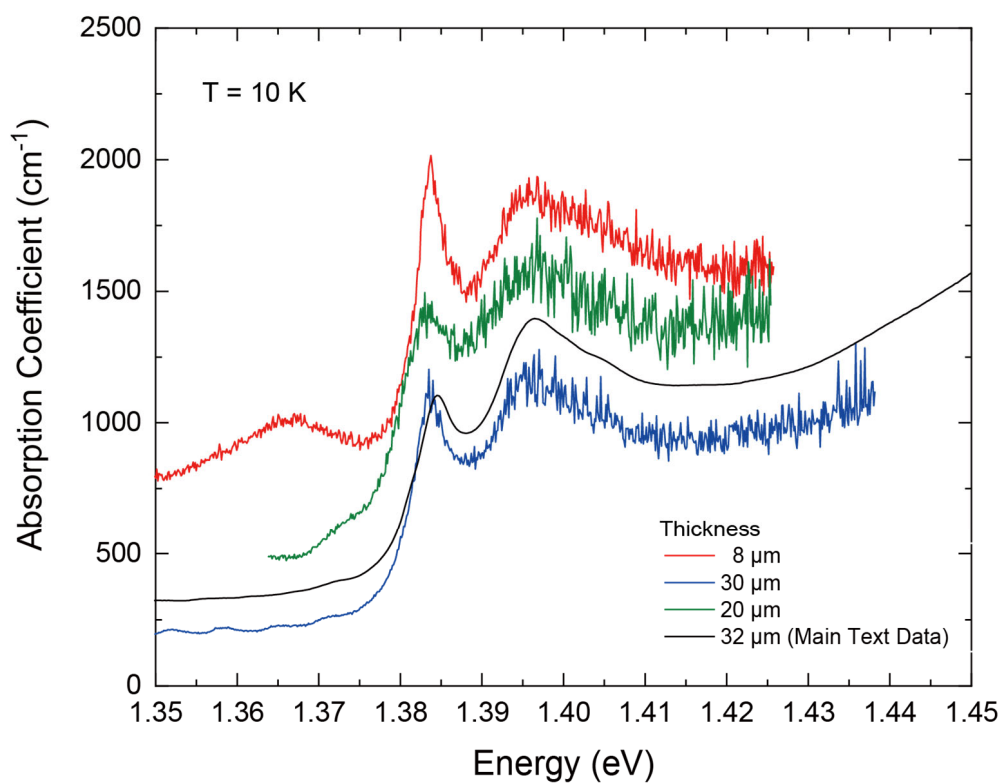

**Figure S2.** Thickness-dependent absorption spectra measured at T = 10 K. All absorption experiments show the exciton at 1.384 eV with the same linewidth. Absorption coefficient calculation is dependent on the thickness of the sample, so there may be an error in absorption coefficient due to an error in the sample thickness. However, the feature of the exciton (like a position, a linewidth) is consistently seen regardless of the thickness of the sample.

**PL measurements.**

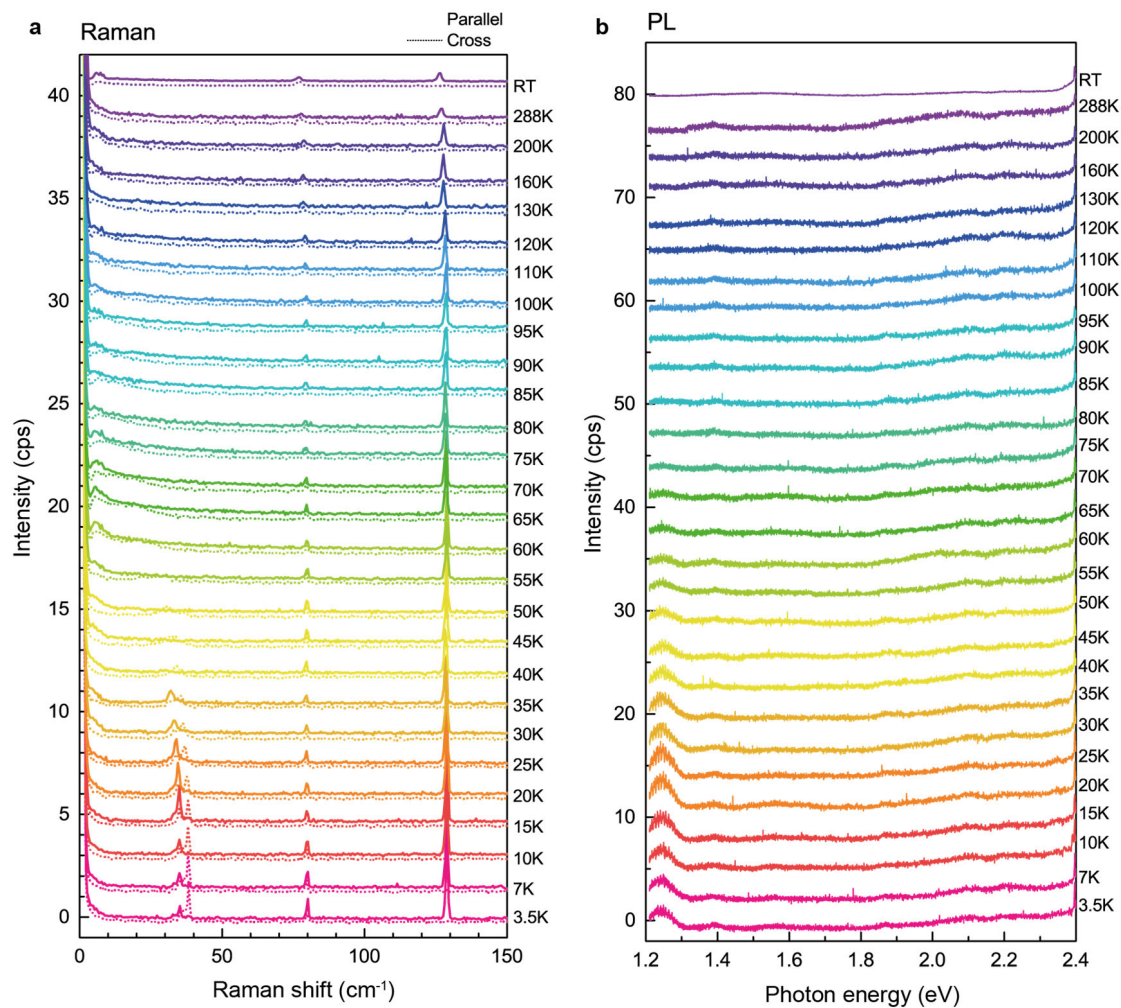

**Figure S3.** Raman and PL spectra of bulk NiI$_2$. a) Raman spectra depicted by solid (dashed) lines are the results of parallel (cross) polarization configuration. b) PL spectra don't have any signal around the position of the absorption peak ~1.38 eV.

**Many-body Calculations.**

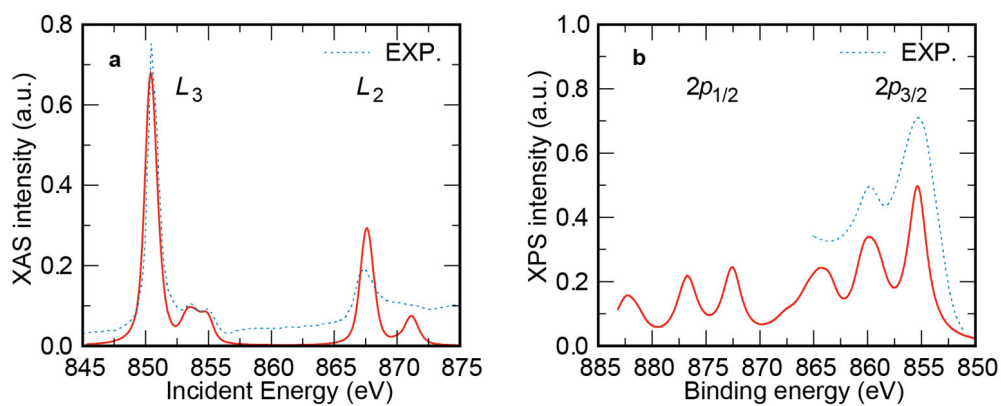

**Figure S4.** Theoretical and experimental data of XAS and XPS. a) Theoretical x-ray absorption spectroscopy (XAS) and b) core-level x-ray photoemission spectroscopy (XPS) spectra of $NiI_2$ cluster calculated with parameters presented in Table 1. Dotted lines in a) and b) refer to the experimental spectra of XAS and XPS, respectively. Experimental XPS spectra were taken after previous literature[1].

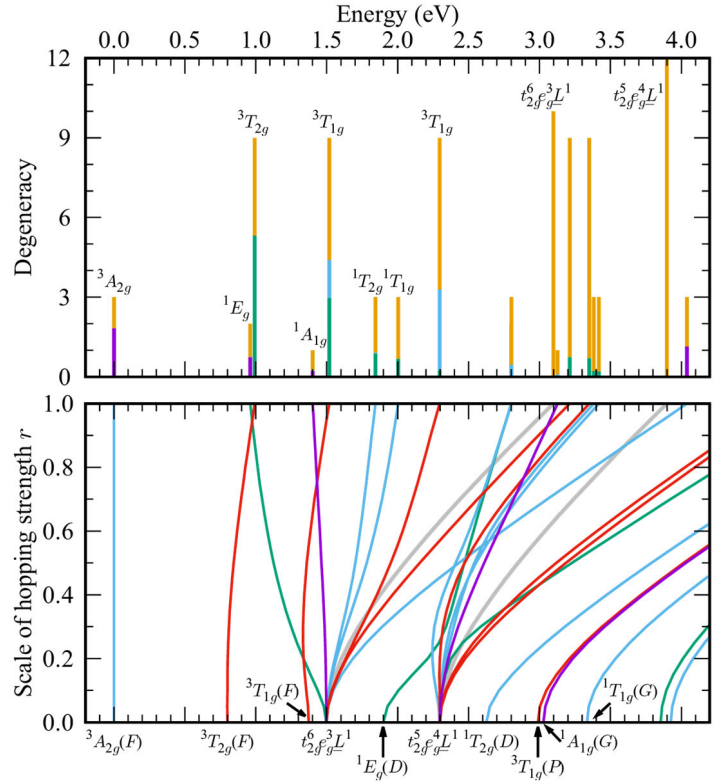

**Figure S5.** Multiplet structure of NiI$_6$ cluster. (top panel) The degeneracy of multiplet levels for $V_{pd\sigma}$ = -0.84 and $V_{pd\pi}$ = 0.5 eV. The length of purple, green, and light blue segments in the top panel refer to the sticks refer to the portion of states with $t_{eg}^6$, $t_{eg}^5 e_g^3$, and $t_{eg}^5 e_g^3$ configurations, respectively, while the length of orange segments refers to remaining states in which electrons are transferred from ligand to Ni site. (bottom panel) The Tanabe-Sugano diagram as functions of hopping strengths characterized by the scale parameter $r$ ($V_{pd\sigma}$ = -0.84 × $r$ and $V_{pd\pi}$ = 0.5 × $r$ eV). Other parameters are given in Table 1.